\documentclass{llncs}
\usepackage{epsfig}
\usepackage[utf8]{inputenc}
\usepackage{multicol}
\usepackage{multirow}
\usepackage{wrapfig}
\usepackage{fancyvrb}

\graphicspath{{./figures/}}

\newcommand{\p}[1]{\texttt{#1}}
\newcommand{\pa}[1]{\p{\{#1\}}}

\begin{document}

\title{Single Time-Stamped Tries for\\Retroactive Call Subsumption}

\author{Flávio Cruz \and Ricardo Rocha}

\institute{CRACS \& INESC-Porto LA, Faculty of Sciences, University of Porto\\
           Rua do Campo Alegre, 1021/1055, 4169-007 Porto, Portugal\\
           \email{\{flavioc,ricroc\}@dcc.fc.up.pt}}

\maketitle


\begin{abstract}
   Tabling is an evaluation strategy for Prolog programs that works by
   storing answers in a table space and then by using them in similar
   subgoals. Some tabling engines use call by subsumption, where it is
   determined that a subgoal will consume answers from a more general
   subgoal in order to reduce the search space and increase
   efficiency. We designed an extension, named Retroactive Call
   Subsumption (RCS), that implements call by subsumption
   independently of the call order, thus allowing a more general
   subgoal to force previous called subgoals to become answer
   consumers. For this extension, we propose a new table space design,
   the Single Time Stamped Trie (STST), that is organized to make
   answer sharing across subsumed/subsuming subgoals simple and
   efficient. In this paper, we present the new STST table space
   design and we discuss the main modifications made to the original
   Time Stamped Tries approach to non-retroactive call by
   subsumption. In experimental results, with programs that stress
   some deficiencies of the new STST design, some overheads may be
   observed, however the results achieved with more realistic programs
   greatly offset these overheads.\\

\textbf{Keywords:} Tabled Evaluation, Call Subsumption, Implementation.
\end{abstract}


\section{Introduction}

Tabling is an evaluation technique for Prolog systems that has several
advantages over traditional SLD resolution: reduction of the search
space, elimination of loops, and better termination
properties~\cite{Chen-96}. In a nutshell, tabling works by storing
found answers in a memory area called the \emph{table space} and then
by reusing those answers in \emph{similar calls} that appear during
the resolution process. First calls to tabled subgoals are considered
\emph{generators} because they are evaluated as usual and their
answers are stored in the table space. Similar calls are named
\emph{consumers} because, instead of executing program code, they are
evaluated by consuming the answers stored in the table space for the
corresponding similar generator subgoal. There are two main approaches
to determine if two subgoals $A$ and $B$ are similar:

\begin{itemize}
\item \emph{Variant-based tabling}: $A$ and $B$ are variants if they
  can be made identical up to variable renaming. For example,
  $p(X,1,Y)$ is a \emph{variant} of $p(W,1,Z)$ because both can be
  transformed into $p(VAR_0,1,VAR_1)$;
\item \emph{Subsumption-based tabling}: $A$ is considered similar to
  $B$ if $A$ is \emph{subsumed} by $B$ (or $B$ \emph{subsumes} $A$),
  i.e., if $A$ is more specific than $B$ (or an instance of).  For
  example, subgoal $p(X,1,f(a,b))$ is subsumed by subgoal $p(Y,1,Z)$
  because there is a substitution $\{Y=X,Z=f(a,b)\}$ that makes
  $p(X,1,f(a,b))$ an instance of $p(Y,1,Z)$. Tabling by call
  subsumption is based on the principle that if $A$ is subsumed by $B$
  and $S_A$ and $S_B$ are the respective answer sets, therefore $S_A
  \subseteq S_B$. Please notice that when using some extra-logical
  features of Prolog, such as the \texttt{var/1} predicate, this
  assumption may not hold.
\end{itemize}

Because subsumption-based tabling can detect a larger number of
similar subgoals, variant and subsumed subgoals, it allows greater
reuse of answers and thus better space usage, since the answer sets
for the subsumed subgoals do not need to be stored. Moreover,
subsumption-based tabling has also the potential to be more efficient
than variant-based tabling because the search space tends to be
reduced as less code is executed~\cite{Johnson-99}. Despite all these
advantages, the more strict semantics of subsumption-based tabling and
the challenge of implementing it efficiently makes variant-based
tabling more popular among the available tabling systems. 

XSB Prolog~\cite{Rao-97} was the first Prolog system providing support
for subsumption-based tabling by introducing a new data structure, the
\emph{Dynamic Threaded Sequential Automata (DTSA)}~\cite{Rao-96}, that
permits incremental retrieval of answers for subsumed
subgoals. However, the DTSA design had one major drawback, namely, the
need for two data structures for the same information. A more space
efficient design, called \textit{Time-Stamped Trie}
(TST)~\cite{Johnson-99,Johnson-00}, solved this by using only one data
structure. Despite the advantages of subsumption-based tabling, the
degree of answer reuse might depend heavily on the call order of the
subgoals. To take effective advantage of subsumption-based tabling in
XSB, the more general subgoals should be called before the specific
ones. When this does not happen, answer reuse does not occur and
Prolog code is executed for both subgoals.

Recently, we implemented a novel design, called \emph{Retroactive Call
  Subsumption (RCS)}~\cite{Cruz-10}, that extends the original TST
approach by also allowing sharing of answers when a subsumed subgoal
is called before a subsuming subgoal. Our extension enables answer
reuse independently of the subgoal call order and therefore increases
the usefulness of subsumption-based tabling. In a nutshell, RCS works
by selectively pruning the evaluation of subsumed subgoals when a more
general subgoal is called and then by restarting the evaluation of the
subsumed subgoal by turning it into a consumer node of the more
general subgoal. To implement RCS we designed the following
components: (i) new control mechanisms for retroactive-based
evaluation; (ii) an algorithm to efficiently retrieve subsumed
subgoals of a subgoal from the table space; and (iii) a new table
space organization, named \emph{Single Time-Stamped Trie (STST)}, that
facilitates the sharing of answers between subsuming/subsumed
subgoals. In this paper, we will focus our discussion on the support
for the STST table space design based on the concrete implementation
we have done on the YapTab tabling engine~\cite{Rocha-00a,Rocha-05a}.

The remainder of the paper is organized as follows. First, we briefly
discuss the background concepts behind tabling and the table space and
we describe how RCS works through an example. Next, we present the
STST design and discuss the main algorithms for answer insertion and
retrieval and how the support data structures are laid out. Then, we
analyze the table space using several benchmarks to stress some
properties of the STST design. Finally, we end by outlining some
conclusions.


\section{Tabling in YapTab}

Tabling is an implementation technique that works by storing answers
from first subgoal calls into the table space so that they can be
reused when a similar subgoal appears. Within this model, the nodes in
the search space are classified as either: \emph{generator nodes}, if
they are being called for the first time; \emph{consumer nodes}, if
they are similar calls; or \emph{interior nodes}, if they are
non-tabled subgoals. In YapTab, we associate a data structure called
\emph{subgoal frame} for each generator node and a data structure
named \emph{dependency frame} for each consumer node. These two data
structures are pushed into two different stacks that are used during
tabled evaluation.


\subsection{Tabling Operations}

In YapTab, programs using tabling are compiled to include
\emph{tabling instructions} that enable the tabling engine to properly
schedule and extend the SLD resolution process. For both variant-based
and subsumption-based tabling, we can synthesize the tabling
instruction set into four main operations:

\begin{description}
\item[Tabled Subgoal Call:] this operation inspects the table space
  looking for a subgoal $S$ similar to the current subgoal $C$ being
  called. For call by variance, we check if a variant subgoal exists,
  and for call by subsumption, we check for subsuming and variant
  subgoals. If a similar subgoal $S$ is found, $C$ will be resolved
  using \emph{answer resolution} and for that it allocates a consumer
  node and starts consuming the set of available answers from $S$. If
  no such $S$ exists, $C$ will be resolved using program clause
  resolution and for that it allocates a generator node and adds a new
  empty entry to the table space.
\item[New Answer:] this operation checks whether a newly found answer
  $A$ for a generator node $C$ is already in the table space. If $A$
  is a repeated answer, the operation fails. Otherwise, $A$ is stored
  as an answer for $C$.
\item[Answer Resolution:] this operation checks whether a consumer
  node $C$ has new answers available for consumption. For call by
  variance, we simply check if new answers are available in the variant
  subgoal $S$, but for call by subsumption, we must determine the new
  relevant answers for $C$ that were stored in the subsuming subgoal
  $S$. If no unconsumed answers are found, $C$ \emph{suspends} and
  execution proceeds according to a specific
  strategy~\cite{Freire-96}. Consumers must suspend because new
  answers may still be found by the corresponding variant/subsuming
  subgoal $S$ that is executing code.
\item[Completion:] this operation determines whether a subgoal $S$ is
  completely evaluated. If this is not the case, this means that there
  are still consumers with unconsumed answers and execution must then
  proceed on those nodes. Otherwise, the operation marks $S$ as
  completed since all answers were found. Future variant or subsumed
  subgoal calls to $S$ can then reuse the answers from the table space
  without the need to suspend.
\end{description}


\subsection{Table Space}

\begin{figure}[ht]
\centering
\includegraphics[width=12cm]{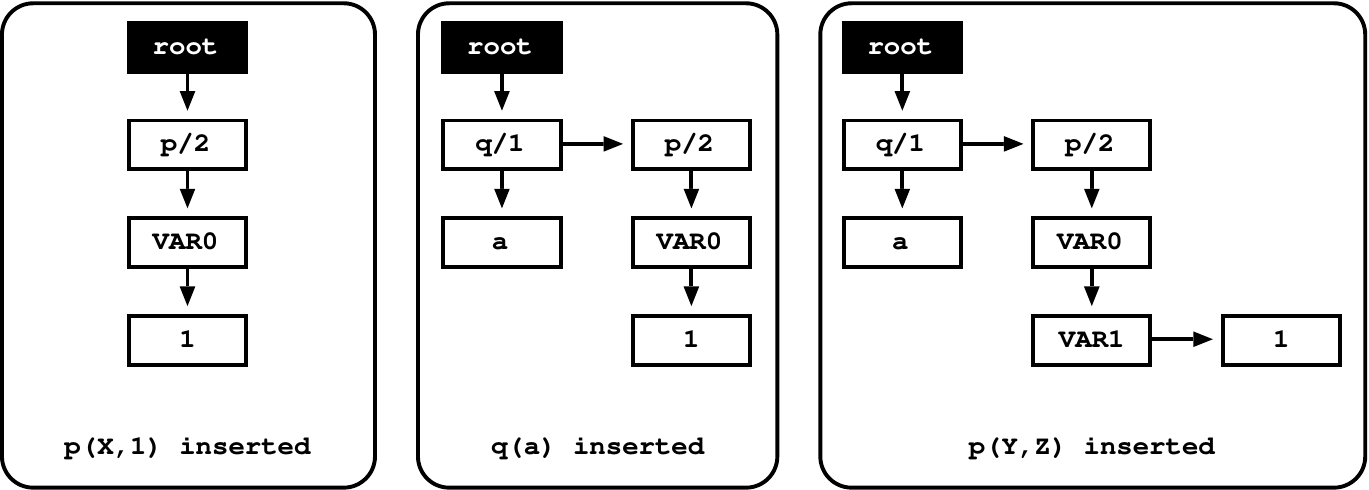}
\caption{Using tries to represent terms}
\label{fig:tries}
\end{figure}

Due to the nature of the previously described tabling operations, the
table space is one of the most important components in a tabling
engine, since the lookup, insertion and retrieval of subgoals and
answers must be done efficiently. Arguably, the most successful data
structure used to implement the table space is
\emph{tries}~\cite{RamakrishnanIV-99}, a tree-like structure where
common prefixes are represented only once. Figure~\ref{fig:tries}
shows an example of using tries to represent terms. In a tabling
setting, tries are used in two levels: in the first level, the
\emph{subgoal tries}, each trie stores the subgoal calls for the
corresponding tabled predicate; in the second level, the \emph{answer
  tries}, each trie stores the answers for the corresponding subgoal
call.

In variant-based tabling, each tabled predicate has a \emph{table
  entry} data structure that contains information about the predicate
and a pointer to the subgoal trie. A trie leaf node in the subgoal
trie corresponds to a unique subgoal call and points to a data
structure called the \emph{subgoal frame}. The subgoal frame contains
information about the subgoal, namely, the state of the subgoal and a
pointer to the corresponding answer trie. In subsumption-based tabling
based on the TST design, we have \emph{subsumptive subgoal frames},
for subgoals that generate answers, and \emph{subsumed subgoal
  frames}, for subgoals that consume answers from subsumptive
subgoals~\cite{Johnson-99}.

\begin{figure}[!ht]
\begin{minipage}{0.5\linewidth}
\centering
\includegraphics[width=5.5cm]{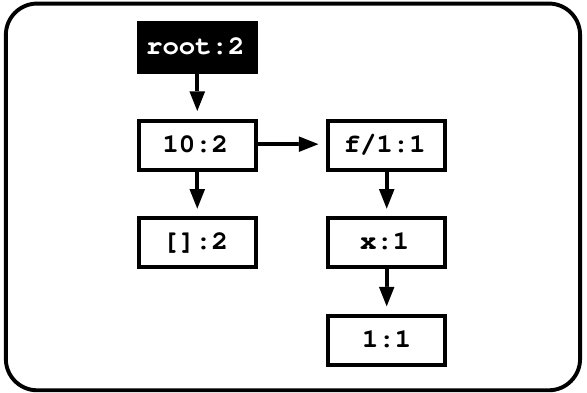}
\caption{A time-stamped answer trie}
\label{fig:tst_base}
\end{minipage}
\begin{minipage}{0.5\linewidth}
\centering
\includegraphics[width=5.5cm]{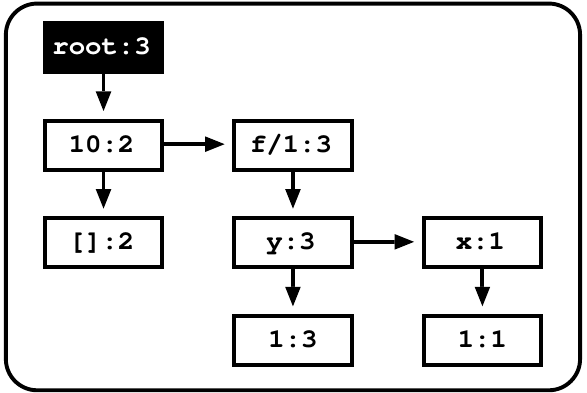}
\caption{Inserting the answer \pa{f(y),1}}
\label{fig:tst_new}
\end{minipage}
\end{figure}

Subsumptive subgoal frames are similar to variant subgoal frames, but
they point to a \emph{time-stamped answer trie} instead, which is an
answer trie where each trie node is extended with \emph{timestamp}
information. Consider, for example, the subgoal call \p{p(VAR0,VAR1)}
and the time-stamped answer trie in Fig.~\ref{fig:tst_base}. The trie
stores 2 answers, \pa{f(x),1} inserted with timestamp 1 and \pa{10,[]}
inserted with timestamp 2. The root node contains the \emph{predicate
  timestamp} and is incremented every time a new answer is
inserted. Consider now that we insert the answer \pa{f(y),1} (see
Fig.~\ref{fig:tst_new}). For this, we increment the predicate
timestamp to 3 and then we set the timestamp of each node on the trie
path of the new answer also to 3. Notice that if we look at leaf nodes
we are able to discern in which order the answers were inserted,
because each new answer is numbered incrementally.

Subsumed subgoal frames store a pointer to the corresponding
subsumptive subgoal frame (the more general subgoal) instead of
pointing to their own answer tries. The frames have a \emph{answer
  return list}, a list of pointers to the relevant answers in the
subsumptive answer trie and a \emph{consumer timestamp} used for
\emph{incremental retrieval} of answers from the subsumptive answer
trie. To consume answers, a subsumed subgoal first traverses its
answer return list checking for more answers, and then executes a
retrieval algorithm in the subsumptive answer trie in order to collect
the answers with newer timestamps, which are then added to the answer
return list. As an example, consider again the answer trie in
Fig.~\ref{fig:tst_new} of the subgoal \p{p(VAR0,VAR1)}. We are now
interested in the incremental retrieval of relevant answers of the
consumer subgoal \p{p(VAR0,1)}. For this, we need to do a depth-first
search on the answer trie using the consumer timestamp as a filter to
ignore already retrieved answers, as we are only interested in answers
that were added after the last retrieval operation. Assuming that the
consumer timestamp was 1, we would retrieve the answer \pa{f(y),1} and
add it to the answer return list to be consumed next.


\section{Retroactive Call Subsumption}

RCS~\cite{Cruz-10} is an extension to subsumption-based evaluation
that enables answer reuse independently of the call order of the
subgoals. While non-retroactive subsumption-based tabling only allows
sharing of answers when a subsumed subgoal is called after a subsuming
subgoal, RCS works around this drawback by selectively pruning the
evaluation of subsumed subgoals and by turning them into consumers.

Let's consider a subgoal $R$ that is subsumed by a subgoal $S$. To do
retroactive evaluation, we must prune the evaluation of $R$, first by
knowing which parts of the execution stacks are involved in its
computation and then by transforming the choice point associated with
$R$ into a consumer node, in such a way that it will consume answers
from the subsuming subgoal $S$, instead of continuing the execution as
a generator. A vital part in this process is that we need to know the
set of answers $A_{old}$, that were already computed by $R$, so that,
when we transform $R$ into a consumer we only consume the set of
answers $A_{new}$, that will be created by $S$. In other words, we
must ensure that the final set of answers $A$ for $R$ is $A = A_{new}
\cup A_{old}$ with $A_{new} \cap A_{old} = \emptyset$. If we do not
obey this principle, the evaluation will not be wrong, but several
execution branches will be executed more than once, thus eliminating
the potential advantage of RCS evaluation.

In RCS, we consider two types of pruning of subgoals. The first type
is \emph{external pruning} and occurs when $S$ is an \emph{external
  subgoal} to the evaluation of $R$. The second one is \emph{internal
  pruning} and occurs when $S$ is an \emph{internal subgoal} to the
evaluation of $R$. Both cases are very similar in terms of the
challenges and problems that arise when doing
pruning~\cite{Cruz-10}. Here, we present an example of external
pruning, that will help us to understand how RCS works, and then we
focus on how the constraint $A_{new} \cap A_{old} = \emptyset$
presented before requires a better table space design than the one
presented in the previous section. Consider thus the query goal
`\p{?-~r(1,X),~r(Y,Z)}' and the following program.

\begin{verbatim}
   :- use_retrosubsumptive_tabling r/2.
   r(1,a).
   r(Y,Z) :- ...
\end{verbatim}

Execution starts by calling \p{r(1,X)}, which creates a new generator
to execute the program code, and a first answer for \p{r(1,X)},
\pa{X=a}, is found. In the continuation, \p{r(Y,Z)} is called, which
will be a subsumptive subgoal for \p{r(1,X)}. Thus, \p{r(1,X)} needs
to be pruned and turned into a consumer of \p{r(Y,Z)}. To prune, we
turn the node of \p{r(1,X)} into a retroactive node that will later be
transformed into either a consumer or a loader node\footnote{A loader
  node works like a consumer node but without suspending the
  computation after consuming the available answers, since the
  corresponding subgoal is completed.}, if \p{r(Y,Z)} does not
complete or completes, respectively. In both cases, when backtracking
to \p{r(1,X)}, we need to consume only the new answers relevant to
\p{r(1,X)} from \p{r(Y,Z)} that were not computed when \p{r(1,X)} was
a generator (in this case, the answer \pa{X=a}).


\section{Single Time Stamped Trie}

Once a pruned subgoal is reactivated and transformed into a consumer
or loader node, it is important to avoid consuming answers that were
found as a generator. In order to efficiently identify such answers,
we designed the \textit{Single Time Stamped Trie (STST)} table space.

\begin{figure}[ht]
\centering
\includegraphics[width=11cm]{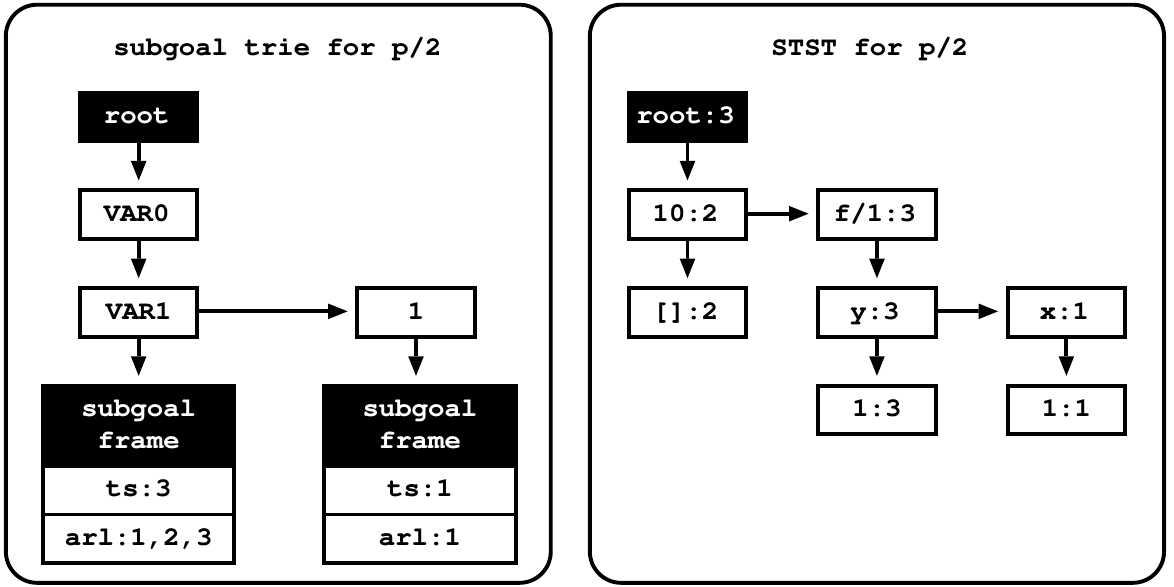}
\caption{The new STST table space organization}
\label{fig:stst}
\end{figure}

In this new organization, each tabled predicate has two tries, the
subgoal trie, as usual, and the STST, a time-stamped answer trie
common to all subgoal calls for the predicate, while each subgoal
frame has an answer return list that references the matching answers
from the STST. Figure~\ref{fig:stst} illustrates an example of the new
table space organization for a tabled predicate \texttt{p/2} with the
subgoals \texttt{p(VAR0,1)} and \texttt{p(VAR0,VAR1)} and the answers
\pa{f(x),1}, \pa{10,[]} and \pa{f(y),1}. This new organization reduces
memory usage, since an answer is represented only once, and permits
easy sharing of answers between subgoals, as the same answer can be
referenced by multiple subgoal frames.

For the subgoal frames, we have also extended them with a \texttt{ts}
field that stores the timestamp of the last generated or consumed
answer. At any time, the answers in the answer return list (field
\texttt{arl} in Fig.~\ref{fig:stst}) are thus the matching answers
from the STST that have a timestamp between 0 and \texttt{ts}. When we
turn a node from generator to consumer, we can collect new answers by
using the timestamp stored in the corresponding subgoal frame, which
was the timestamp of the last answer successfully inserted into the
STST.

When a subsumed subgoal is pruned, we know its timestamp and we can
easily turn it into a consumer, since now instead of inserting
answers, the new consumer will now consume them from the STST, like in
the TST design, by incrementally retrieving answers from
it. Therefore, the cost of such transformation is very low given that
both generators and consumers use an answer return list and a
timestamp. If we had used the original TST design, the pruned subgoal
would have its own answer trie, call it $T_1$, and we would need to,
before consuming answers, check if the answers on $T_1$ have already
appeared on the answer trie of the subsuming subgoal, call it $T_2$.
Such a task is quite complex, since answers in $T_1$ are instances of
the answers in $T_2$.

Notice that in both variant-based and subsumption-based tabling, only
the substitutions for the variables in a subgoal call are stored in
the answer tries~\cite{RamakrishnanIV-99}. For example, for the
subgoal \p{p(VAR0,1)} and the answers \pa{f(x),1} and \pa{f(y),1},
only the substitutions \pa{f(x)} and \pa{f(y)} are stored, since
during consumption of answers only the substitutions are used for
unification. However, in the STST design, we cannot do this, since any
subgoal of \p{p/2} can use the answers stored in the answer trie,
therefore we need to store all the subterms of each answer.


\subsection{Inserting Answers}

The insertion of answers in the STST works like the insertion of
answers in standard TSTs, but special care must be taken when updating
the \texttt{ts} field on the subgoal frames. When only one subgoal is
adding answers to the STST, the \texttt{ts} field is incremented each
time an answer is inserted. Repeated answers are easily recognized by
testing if the answer is new or not by using the \texttt{ts}
field. The problem arises when several subgoals are inserting answers,
as it may be difficult to determine when an answer is new or repeated
for a certain subgoal.

Let's consider two subgoals of the same predicate $p$, $S_1$ and
$S_2$, and their corresponding timestamps, $T_1$ and $T_2$. $S_1$ has
found and inserted the first 3 answers ($T_1 = 3$) in the STST and
$S_2$ then started evaluating and inserted the next 3 answers, answers
4, 5 and 6 ($T_{2} = 6$). Now, when execution backtracks to $S_1$,
answer 5 is found and, while it is already on the trie, it must be
considered as a new answer for $S_1$.

By default, we could consider answer 5 as new, since $T_1$ is in the
past ($T_1 < 5$).  But this can also lead to problems if next we
update $T_1$ to either 6 (the predicate timestamp) or 5 (the timestamp
for answer 5). For example, if later, answer 4 is also found for
$S_1$, it will be considered as a repeated answer during its insertion
since now $T_1 > 4$. Therefore, we need a more complex mechanism to
detect repeated subgoal answers.

In our approach, we use a \emph{pending answer index} for each subgoal
frame. This index contains all the answers that are older than the
current subgoal frame timestamp field but that have not yet been found
by the subgoal. It is built whenever the timestamp of the answer being
inserted is younger than the subgoal frame timestamp, by collecting
all the relevant answers in the STST with a timestamp younger than the
current subgoal frame timestamp. Later, when an answer is found but is
already on the trie, and therefore will have an older timestamp than
the subgoal frame timestamp, we must lookup on the pending answer
index to check if the answer is there. If so, we consider it a new
answer and remove it from the index; if not, we consider it a repeated
answer.

The pending answer index is implemented as a single linked list, but
can be transformed into a hash table if the list reaches a certain
threshold. In Fig.~\ref{fig:stst_insert_answer}, we present the code
for the \texttt{stst\_insert\_answer()} procedure, which given an
answer and a subgoal frame, inserts the answer into the corresponding
STST for the subgoal frame. The pseudo-code is organized into four
cases:

\begin{enumerate}
\item Answers are inserted in order by the same subgoal. This is the
  most common situation.
\item The answer being inserted is the only answer in the STST that
  the current subgoal has still not considered. It is trivially marked
  as a new answer.
\item The timestamp of the answer being inserted is older than the
  subgoal frame timestamp. The pending answer index must be
  consulted.
\item The timestamp of the answer being inserted is younger than the
  subgoal frame timestamp $t$. We must collect all the relevant
  answers in the STST with a timestamp younger than $t$ (calling
  \texttt{collect\_relevant\_answers()}) and add them
  to the pending answer index, except for the current answer.
\end{enumerate}

\begin{figure}[!ht]
\begin{verbatim}
stst_insert_answer(answer, sg_fr) {
  table_entry = table_entry(sg_fr)
  stst = answer_trie(table_entry)
  old_ts = predicate_timestamp(stst)
  leaf_node = answer_check_insert(answer, stst)
  leaf_ts = timestamp(leaf_node)
  new_ts = predicate_timestamp(stst)

  if (new_ts == old_ts + 1 and ts(sg_fr) == old_ts)
    // case 1: incremental answer by the same subgoal
    ts(sg_fr) = new_ts
    return leaf_node
  else if (new_ts == old_ts == leaf_ts and ts(sg_fr) == new_ts - 1)
    // case 2: only answer still not considered by the current subgoal
    ts(sg_fr) == new_ts
    return leaf_node
  else if (leaf_ts <= ts(sg_fr))
    // case 3: answer with a past timestamp, check pending answer index
    if (is_in_pending_answer_index(leaf_node, sg_fr))
      remove_from_pending_answer_index(leaf_node, sg_fr)
      return leaf_node
    else
      return NULL
  else
    // case 4: answers were inserted by someone else
    ans_tpl = answer_template(sg_fr)
    pending_list = collect_relevant_answers(ts(sg_fr),ans_tpl,stst)
    remove_from_list(leaf_node, pending_list)
    add_to_pending_answer_index(pending_list, sg_fr)
    ts(sg_fr) = new_ts
    return leaf_node
}
\end{verbatim}
\caption{Pseudo-code for procedure \texttt{stst\_insert\_answer()}}
\label{fig:stst_insert_answer}
\end{figure}

Note that when a generator subgoal frame is transformed into a
consumer subgoal frame, we remove all the answers from the pending
answer index and we safely insert them on the answer return list. With
this, all the consumer mechanisms can be used as usual.


\subsection{Reusing Answers}

The STST approach also allows reusing answers when a new subgoal is
called. As an example, consider that two unrelated (no subsumption
involved) subgoals $S_1$ and $S_2$ are fully evaluated. If a subgoal
$S$ is then called, it is possible that some of the answers on the
STST match $S$ even if $S$ neither subsumes $S_1$ or $S_2$. Hence,
instead of eagerly running the predicate clauses, we can start by
loading the matching answers already on the STST, which can be enough
if, for example, $S$ is pruned by a cut. This is a similar approach
to the \textit{incomplete tabling} technique for variant-based
tabling~\cite{Rocha-06a}.

While the reuse of answers has some advantages, it can also lead to
redundant computations. This happens when the evaluation of $S$
generates more general answers than the ones initially stored on the
STST. For an example, consider the retroactive tabled predicate
\p{p/2} with only one fact, \p{p(X,a)}. If subgoal \p{p(1,Y)} is first
called, the answer represented as \pa{1,a} is added to the STST for
\p{p/2} and execution would succeed. If the subgoal \p{p(X,Y)} is then
called, we would search the STST for relevant answers and the first
answer would be \pa{1,a}. If we ask for more answers, the system would
return a new answer, \pa{VAR0,a}, and add it to the STST. On the other
hand, if we called \p{p(X,Y)} with an empty STST, only the answer
\pa{VAR0,a} would be returned.


\subsection{Answer Templates}

The \emph{answer template} is a data structure that is built on the
choice point stack and is pushed into the stack when a new subgoal,
generator or consumer, is called. The contents of the answer template
are the terms from the subgoal call that must be read when inserting a
new answer, if a generator, or the terms from the subgoal call that
must be unified when consuming answers, if a consumer.

On variant-based tabling, the answer template is just the set of
variables found in the subgoal call, since we only store variable
substitutions on the answer trie. For non-retroactive call by
subsumption, where we use an answer trie per generator subgoal, the
answer template for each consumer subgoal is built according to its
generator subgoal. For example, if the subsumptive subgoal is
\p{p(1,f(X),Y)} and the subsumed subgoal is \p{p(1,f([A,B]),a(C))},
the answer template for the subsumed subgoal will be
\pa{[A,B],a(C)}.

With RCS, the answer template is simply built by copying the full set
of argument registers for the generator or consumer call. This is a
very efficient operation compared to non-retroactive call by
subsumption. Notice that we need the full answer template because the
answers stored on the STST contain all the predicate arguments, hence
the unification of matching answers must be seen as unifying against
the most general subgoal.


\subsection{Compiled Tries}

Compiled tries are a well-known implementation mechanism in which we
decorate a trie with WAM instructions when a subgoal completes in such
a way that, instead of consuming answers one by one in a bottom-up
fashion, we execute the trie instructions in order to consume answers
incrementally in a top-down fashion, thus taking advantage of the
nature of tries~\cite{RamakrishnanIV-99}.

Our approach only compiles the STST when the most general subgoal is
completed. This avoids problems when a subgoal is executing compiled
code and another subgoal is inserting answers, leading to the loss of
answers as hash tables can be dynamically created and expanded.

With this optimization, we can throw away the subgoal trie and the
subgoal frames when the most general subgoal completes and the STST is
compiled. Later, when a new subgoal call is made, we just build the
answer template by copying the argument registers and then we execute
the compiled trie, thus bypassing all the mechanisms of locating the
subgoal on the subgoal trie, leading to memory and speedup gains.


\section{Experimental Results}

Previous experiments using the STST design for comparing RCS with
non-retroactive call by subsumption, showed good results when
executing programs that take advantage of RCS~\cite{Cruz-10}. Despite
these good results and while the STST presents some conveniences in
implementing RCS, here we will focus in measuring the impact of having
to store the complete answers on the STST, instead of storing only the
variable substitutions, since it is more expensive to insert/load
terms to/from the STST. To this purpose, we measured the overheads of
several programs that stress this weakness, both in terms of time and
space.

The environment for our experiments was a PC with a 2.66 GHz Intel
Core(TM) 2 Quad CPU and 4 GBytes of memory running the Linux kernel
2.6.31 with YapTab 6.03. For benchmarking, we used six different
versions of the well-known \texttt{path/2} program, that computes the
reachability between nodes in a graph, with several dataset
configurations: \texttt{chain}, where each node $i$ connects with node
$i+1$; \texttt{cycle}, like a chain, but the last node connects with
the first; \texttt{grid}, where nodes are organized in a square
configuration; \texttt{pyramid}, where nodes form a pyramid; and
\texttt{tree}, a binary tree.  To increase the size of the terms to be
stored in the STST, we transformed, both the programs and datasets, to
use a functor term in each argument, instead of simple integers. For
example, a fact \texttt{edge(3,4)} was transformed into
\texttt{edge(f(3),f(4))} and the transformed version of the
\textbf{left\_first} (left recursion with the recursive clause first)
\texttt{path/2} program is:

\begin{verbatim}
   path(f(X),f(Z)) :- path(f(X),f(Y)), edge(f(Y),f(Z)).
   path(f(X),f(Z)) :- edge(f(X),f(Z)).
\end{verbatim}

We experimented the six different versions of the \texttt{path/2}
program with different graph sizes for the datasets using a query
goal, \texttt{path(f(X),f(Y))}, that does not take advantage of RCS
evaluation, i.e., never calls more general subgoals after specific
ones.


\subsection{Execution Times}

In Table~\ref{tbl:stst_times}, we present the execution times, in
milliseconds, for RCS evaluation and the respective overheads for
variant-based and non-retroactive subsumption-based tabling. Each
execution time is the average of 3 runs.

\begin{table}[!ht]
\caption{Execution times, in milliseconds, for RCS evaluation and the
  respective overheads for variant-based and non-retroactive
  subsumption-based tabling for the query goal
  \texttt{path(f(X),f(Y))} in YapTab}
\centering
\begin{tabular}{llrcc}
\hline\hline
\multicolumn{2}{c}{\multirow{2}{*}{\textbf{Program/Dataset}}} & 
\multicolumn{3}{c}{\textbf{YapTab}} \\ \cline{3-5}
& & \textbf{~RCS~} & \textbf{~~~RCS/Var~~~} & \textbf{RCS/Sub} \\
\hline
\multirow{5}{*}{\textbf{left\_first}}
& \textbf{chain (2048)}   &    812 & 1.21 & 1.30 \\
& \textbf{cycle (2048)}   &  1,722 & 1.08 & 1.14 \\
& \textbf{grid (64)}      & 17,261 & 1.38 & 1.17 \\
& \textbf{pyramid (1024)} &    869 & 1.18 & 1.23 \\
& \textbf{tree (65536)}   &    573 & 1.23 & 1.17 \\
& \textit{Average}        &        & 1.22 & 1.20 \\
\hline
\multirow{5}{*}{\textbf{left\_last}}
& \textbf{chain (2048)}   &    894 & 1.35 & 1.42 \\
& \textbf{cycle (2048)}   &  1,794 & 1.13 & 1.29 \\
& \textbf{grid (64)}      & 18,187 & 1.47 & 1.23 \\
& \textbf{pyramid (1024)} &    862 & 1.14 & 1.24 \\
& \textbf{tree (65536)}   &    582 & 1.27 & 1.19 \\
& \textit{Average}        &        & 1.27 & 1.27 \\
\hline
\multirow{5}{*}{\textbf{right\_first}}
& \textbf{chain (4096)}   &  4,004 & 1.21 & 1.21 \\
& \textbf{cycle (4096)}   &  7,324 & 1.09 & 1.04 \\
& \textbf{grid (32)}      &  1,130 & 1.27 & 1.29 \\
& \textbf{pyramid (2048)} &  3,172 & 1.17 & 1.10 \\
& \textbf{tree (32768)}   &    300 & 1.15 & 1.04 \\
& \textit{Average}        &        & 1.18 & 1.14 \\
\hline
\multirow{5}{*}{\textbf{right\_last}}
& \textbf{chain (4096)}   &  3,642 & \textbf{0.95} & 1.08 \\
& \textbf{cycle (4096)}   &  7,965 & \textbf{0.98} & 1.12 \\
& \textbf{grid (32)}      &    997 & \textbf{0.96} & 1.22 \\
& \textbf{pyramid (2048)} &  3,170 &         1.09  & 1.12 \\
& \textbf{tree (32768)}   &    528 &         1.76  & 1.05 \\
& \textit{Average}        &        &         1.15  & 1.12 \\
\hline
\multirow{5}{*}{\textbf{double\_first}}
& \textbf{chain (256)}    &  1,708 & 3.94 & 1.58 \\
& \textbf{cycle (256)}    &  2,945 & 1.11 & 1.51 \\
& \textbf{grid (16)}      &  3,936 & 1.53 & 1.53 \\
& \textbf{pyramid (256)}  &  7,480 & 4.04 & 1.44 \\
& \textbf{tree (16384)}   &    766 & 2.19 & 1.35 \\
& \textit{Average}        &        & 2.56 & 1.48 \\
\hline
\multirow{5}{*}{\textbf{double\_last~~~}}
& \textbf{chain (256)}    &  1,778 & 4.08 & 1.69 \\
& \textbf{cycle (256)}    &  2,932 & 1.10 & 1.64 \\
& \textbf{grid (16)}      &  3,956 & 1.57 & 1.54 \\
& \textbf{pyramid (256)}  &  6,829 & 3.70 & 1.30 \\
& \textbf{tree (16384)}   &    720 & 2.04 & 1.29 \\
& \textit{Average}        &        & 2.50 & 1.49 \\
\hline
& \textit{Total Average}  &        & 1.65 & 1.28 \\
\hline\hline
\end{tabular}
\label{tbl:stst_times}
\end{table}

From these results, we can observe that, on total average for these
set of benchmarks, the transformed \texttt{path/2} program has an
overhead of 65\% and 28\% when compared with variant-based and
non-retroactive subsumption-based tabling, respectively. The insertion
of new answers into the table space and the consumption of answers are
the primary causes for these overheads. The programs with the worst
overheads are \textbf{double\_first} and \textbf{double\_last}, with
48\% and 49\% of overhead against non-retroactive subsumption-based
tabling. These programs also create the higher number of consumers,
both variant consumers and subsumed consumers than any other benchmark
in these experiments. The \textbf{right\_first} and
\textbf{right\_last} only create subsumed consumers, and they have an
overhead of 14\% and 12\%, respectively, against non-retroactive call
by subsumption, which are the lowest overhead values. In the
\textbf{left\_first} and \textbf{left\_last} programs, only one
variant consumer is allocated, however, on average, they perform worse
than the \textbf{right} versions.

We thus argue that the number of consumer nodes can greatly reduce the
applicability and performance of the STST table space organization
when the operation of loading an answer from the trie is more
expensive. While this situation seems disadvantageous, execution time
can be reduced if another subgoal call appears (for example
\texttt{path(X,Y)}) where it is possible to reuse the answers from the
table before executing the predicate clauses.


\subsection{Memory Usage}

We executed the previous benchmarks and measured the number of answer
trie nodes stored for each program. Table~\ref{tbl:stst_space}
presents such numbers for RCS evaluation and the relative numbers for
variant and subsumption-based tabling. The programs
\textbf{left}, \textbf{right} and \textbf{double} are the left, right
and double versions of the \texttt{path/2} program (note that the
number of stored trie nodes is the same for both the versions of the
\texttt{path/2} program, with the recursive clause first or with the
recursive clause last).

\begin{table}[!ht]
\caption{Number of stored answer trie nodes for RCS evaluation and the
  relative numbers for variant-based and non-retroactive
  subsumption-based tabling for the query goal
  \texttt{path(f(X),f(Y))} in YapTab}
\centering
\begin{tabular}{llrcc}
\hline\hline
\multicolumn{2}{c}{\multirow{2}{*}{\textbf{Program/Dataset}}} & 
\multicolumn{3}{c}{\textbf{YapTab}} \\ \cline{3-5}
& & \textbf{~~\#RCS~~} & \textbf{~~~Var/RCS~~~} & \textbf{Sub/RCS} \\
\hline
\multirow{5}{*}{\textbf{left}}
& \textbf{chain (2048)}   &  2,100,233 & 0.99902 & 0.99902 \\
& \textbf{cycle (2048)}   &  4,200,450 & 0.99902 & 0.99902 \\
& \textbf{grid (64)}      & 16,789,506 & 0.99951 & 0.99951 \\
& \textbf{pyramid (1024)} &  1,576,457 & 0.99870 & 0.99870 \\
& \textbf{tree (65536)}   &    983,056 & 0.96665 & 0.96665 \\
& \textit{Average}        &            & 0.99258 & 0.99258 \\
\hline
\multirow{5}{*}{\textbf{right}}
& \textbf{chain (4096)}   &  8,398,847 & 1.99756 & 0.99902 \\
& \textbf{cycle (4096)}   & 16,789,506 & 1.99902 & 0.99951 \\
& \textbf{grid (32)}      &  1,051,650 & 1.99610 & 0.99805 \\
& \textbf{pyramid (2048)} &  6,302,719 & 1.99675 & 0.99870 \\
& \textbf{tree (32768)}   &    491,520 & 1.76667 & 0.90000 \\
& \textit{Average}        &            & 1.95122 & 0.97906 \\
\hline
\multirow{5}{*}{\textbf{double~~~}}
& \textbf{chain (256)}    &     26,387 & 2.48365 & 0.87490 \\
& \textbf{cycle (256)}    &     36,844 & 3.57141 & 0.89333 \\
& \textbf{grid (16)}      &     59,028 & 2.22920 & 0.82586 \\
& \textbf{pyramid (256)}  &     56,638 & 3.47583 & 0.95187 \\
& \textbf{tree (16384)}   &    213,008 & 1.88449 & 0.96148 \\
& \textit{Average}        &            & 2.72892 & 0.90149 \\
\hline
& \textit{Total Average}  &            & 1.89091 & 0.95771 \\
\hline\hline
\end{tabular}
\label{tbl:stst_space}
\end{table}

From these results we can observe that, on total average for this set
of benchmarks, the variant-based table design requires 1.89 times more
memory space than the STST table space organization. In particular,
for the \textbf{double} program, these differences are higher because
in the variant-based design there are more generator subgoal calls and
thus more answer tries are created.

When comparing RCS to the non-retroactive subsumption-based engine,
the latter only stores, on total average for this set of benchmarks,
around 4\% less trie nodes than RCS evaluation, even if the
\texttt{f/1} functor terms need to be stored in the STST. This is
easily understandable because the first \texttt{f/1} functor term is
only represented once, at the top of the STST, and then there is one
second \texttt{f/1} functor for each node in the graph, therefore, the
total number of functors stored in the STST is insignificant when
compared to the total number of terms stored in the trie. Also note
that, for the \textbf{double} benchmarks, the datasets used are small
if compared to the datasets used for the other benchmarks, but the
space overhead is more significant (18\% in the worst case). We thus
argue that the cost of the extra space needed to store terms in the
STST is less significant as more terms are stored in the tries.


\section{Conclusions}

We presented a new table space organization that is particularly well
suited to be used with Retroactive Call Subsumption. Our proposal uses
ideas from the original TST design and innovates by having only a
single answer trie per predicate, making it easier to share answers
across subgoals for the same predicate. We presented the challenges
when using a single answer trie and how they have been solved, for
example, with the use of pending answer indices. Moreover, we think
that the new design should not be very difficult to port to other
tabling engines, since it uses the trie data structure extended with
timestamp information.

Our experiments with RCS showed promising results when used with
programs that take advantages of the new mechanisms. In this paper, we
benchmarked and discussed the overhead in terms of time and space when
storing and loading complete answers, instead of using variable
substitutions, for programs that do not takes advantage of RCS
evaluation, i.e., that never call more general subgoals after specific
ones. Our results show that the time overhead can be noticeable,
however, in terms of space, the number of extra trie nodes appears to
be low.


\section*{Acknowledgments}

This work has been partially supported by the FCT research projects
HORUS (PTDC/EIA-EIA/100897/2008) and LEAP (PTDC/EIA-CCO/112158/2009).


\bibliographystyle{splncs}
\bibliography{references}

\end{document}